\documentclass{article}

\usepackage[english]{babel}

\usepackage[letterpaper,top=2cm,bottom=2cm,left=3cm,right=3cm,marginparwidth=1.75cm]{geometry}

\usepackage{amsmath}
\usepackage{graphicx}
\usepackage[colorlinks=true, allcolors=blue]{hyperref}
\usepackage[T1]{fontenc}
\usepackage[utf8]{inputenc}
\usepackage{authblk}

\title{Structural lubricity of physisorbed gold clusters on graphite and its breakdown: Role of boundary conditions and contact lines}

\author[1]{Hongyu Gao}
\author[1]{Martin H. Müser\thanks{martin.mueser@mx.uni-saarland.de}}

\affil[1]{Department of Materials Science and Engineering, Saarland University, 66123 Saarbr\"ucken, Germany}

\providecommand{\keywords}[1]{\textbf{\textit{Keywords--}} #1}

\begin{document}
\maketitle

\begin{abstract}

The sliding motion of gold slabs adsorbed on a graphite substrate is simulated using molecular-dynamics.
The central quantity of interest is the mean lateral force, i.e., the kinetic friction rather than the maximum lateral forces, which correlate with the static friction.
For most set-ups, we find Stokesian damping to resist sliding.
However, velocity-insensitive (Coulomb) friction is observed for finite-width slabs sliding parallel to the armchair direction if the bottommost layer of the three graphite layers is kept at zero stress rather than at zero displacement.
Although the resulting kinetic friction remain much below the noise produced by the erratic fluctuations of (conservative) forces typical for structurally lubric contacts, the nature of the instabilities leading to Coulomb friction could be characterized as quasi-discontinuous dynamics of the Moiré patterns formed by the normal displacements near a propagating contact line.
It appears that the interaction of the graphite with the second gold layer is responsible for the symmetry breaking occurring in the interface when a contact line moves parallel to the armchair rather than to the zigzag direction. 
\end{abstract}


\section{Introduction}

The origin of friction between surfaces in relative sliding motion has lead to speculations since the times of Coulomb \cite{Coulomb1785}.
He recognized as possible friction mechanism that surface asperities must deform so that they can slide past each other or that surface atoms, due to their proximity, can form a coherence, which needs to be overcome to initiate sliding.
In an extreme point of view, individual atoms could be seen as the smallest possible asperities and the molecular generation of coherence as an (incommensurate) superstructure.
When irregularities of a counterbody makes surface atoms deflect from their ideal lattice positions, vibrations are excited.
As long as they can be treated as a small perturbation of the crystal lattice, the ensuing dissipation can be approximately described within linear-response theory~\cite{Adelman1976JCP}, i.e., friction would be Stokes like. 
However, once the system is driven to an instability point, where, at a given moment of time, the closest stable atomic position is a finite distance away from the previous one, an atom -- or a collection of atoms -- quasi-discontinuously pops toward the next available equilibrium site~\cite{Prandtl28}.
As already demonstrated by Prandtl himself, such a process leads to Coulomb like friction, unless the system is thermal and driven so slowly that the relevant degree of freedom can sample both the old and the new potential energy minima even before the ultimate instability point, in which case Stokes friction can arise again.
Interestingly, the transition between the Coulomb and the Stokes regime in the Prandtl model, which consists of a simple damped spring pulled past a sinusoidal potential, can be well described by the Carreau-Yasuda equation, which Yasuda \cite{Yasuda1979}  derived to describe the shear thinning of polystyrene~\cite{Muser2020L}.

Hirano and Shinjo \cite{Hirano1990PRB} were arguably the first to come to the conclusion that ideally flat but incommensurate (iron) solids placed on top of each other do not interlock to a degree leading to static friction, implying the absence of kinetic friction at (infinitesimally) small sliding velocity.
They later called the phenomenon superlubricity~\cite{Shinjo1993SS}.
Their prediction of superlow friction was confirmed experimentally on MoS$_2$ coated solids in ultra-high vacuum:
Martin et al. \cite{Martin1993PRB} measured
friction coefficients as small as $10^{-3}$.
In theory, the small friction originates from the systematic annihilation of lateral forces, where some interfacial atoms are pushed to the right but others to the left. 
Thus, the more incommensurate the interface, the smaller the friction.
This expectation was confirmed in experiments by Dienwiebel et al. \cite{Dienwiebel2004PRL}, who rotated two graphite flakes against each other.
Friction reduces dramatically with increasing angular misfit~\cite{Filippov2008PRL}.
To distinguish ultra-small friction between two (incommensurate) solids in direct mechanical contact from that due to an intervening fluid layer, M\"user~\cite{Muser2004EPL} introduced the term structural lubricity, which can be seen as a special form of superlubricity~\cite{Erdemir2006JPD,Baykara2018APR}.

If either the substrate or the physisorbed cluster is amorphous, simple scaling arguments based on the law of large numbers suggest root-mean square lateral or friction forces $F$ to increase with $\sqrt{A}$ rather than linearly with the area of the cluster's basal plane $A$, as long as elasticity keeps the upper hand over interfacial forces~\cite{Muser2001PRL, Muser2004EPL}.
For incommensurate surfaces, the increase of friction with contact area is generally weaker and may even depend sensitively on the cluster's shape as well as on the sliding direction~\cite{Wijn2012PRB}.
However, in certain situations, even incommensurate structures can yield (maximum) lateral forces that scale  with  $\sqrt{A}$, e.g., when the cluster has a sharp edge that is perfectly aligned with the substrate and the sliding direction is normal to that edge.
For a random contact line, Koren and Duerig~\cite{Koren2016PRB} expected $F \propto {\sqrt[4]{A}}$ from  scaling arguments, since annihilation of lateral forces in incommensurate contacts is quite systematic and randomness comes from the outer rim only, while spherical adsorbed solids (noble-gas monolayers) reveal yet another scaling with approximately $A^{0.37}$~\cite{Varini2015Nanoscale}.
Experiments on metal cluster on graphite  \cite{Dietzel2013PRL,Dietzel2018N,Hartmuth2019L} as well as computer simulations~\cite{Chen2020F} seem to confirm the scaling arguments. 
M\"user et al.~\cite{Muser2001PRL} showed that the scaling arguments can also be motivated from a continuum description of the repulsive forces between interacting surfaces. 
Marom et al.~\cite{Marom2010PRL} introduced a similar parameter for discrete systems, which was called the registry index.
Hod~\cite{Hod2013CPC} demonstrated that it provides quantitative descriptions for various layered compounds.

The just-mentioned scaling arguments assume the solid bodies to be rigid. 
Thus, an important question to be addressed is what makes two crystals with plane surfaces pin and/or what controls kinetic friction.
Analysis of low-dimensional models, such as the Frenkel-Kontorova model, can only provide qualitative insight, as they ignore the long-range nature of elastic restoring forces~\cite{Shinjo1993SS,Muser2004EPL}. 
Candidates to foster pinning between two incommensurate solids with flat surfaces are adsorbed layers or boundary lubricants~\cite{He1999S,Muser2001PRL,Dietzel2008PRL,Feldmann2014PRL} and (extended) lattice defects.
For adsorbed layers to act as pinning agents, normal pressures have to be large enough, since the interfacial layer would simply form an intervening viscous medium between the surfaces otherwise, i.e., depending on the ratio of energy activation barriers and thermal energy, the response of a molecular layer can range from Stokes to Coulomb like friction, as discussed, for example by M\"user~\cite{Muser2020L}.
In fact, \"Ozo$\check{\rm g}$ul et al.~\cite{Ozogul2017APL} found superlubric states for metal clusters adsorbed on graphite despite conducting their experiments under ambient conditions. 
Roughness can also lead to pinning, e.g., through many small contact patches that carry relatively little normal load but, due to their small size, exert relatively large frictional shear stresses~\cite{Muser2019FME}.
Finally, when interfacial interactions dominate the ones inside the bulk, which happens when chemical bonds form across the interface~\cite{Dietzel2017acsNano}, two solids or clusters have no choice but to pin. 

In this article, we will focus on UHV conditions and idealized systems with perfectly smooth surfaces.
Pinning, or rather the onset of pinning in such clean conditions is a competition between elastic restoring forces and interfacial interactions.
The question to be addressed then is which force ``wins'' at large length scales. 
For three-dimensional amorphous systems, M\"user~\cite{Muser2004EPL} predicted elastic restoring forces to scale linearly with the linear dimension of a contact so that details determine which of the two effect keeps the upper hand, since lateral, interfacial forces frequently obey the same scaling.
This prediction is consistent with the results by Sharp et al.~\cite{Sharp2016PRB} and Monti and Robbins~\cite{Monti2020N}, who found that large, disordered clusters pin while small ones don't. 
Also Wang et al.'s~\cite{Wang2019NL} results that the friction of thin graphite sheets crosses over from a $F \propto \sqrt{A}$ to $F \propto A$ scaling are consistent with the scaling arguments, since the flakes are two-dimensional so that a cross-over to linear scaling is expected.

Despite much progress on the presence or rather absence of static friction and the absence of instabilities leading to Coulomb-like friction, it is surprisingly unexplored what parameters affect the prefactors to viscous-like damping in a contact satisfying the requirements for structural lubricity.
In recent work, Lodge
et al.~\cite{Lodge2016SR} compared simulations to experimental results on the slip time of gold nanocrystals sliding past graphene substrate using the quartz crystal microbalance.
However, their work does not reveal how damping depends on the velocity or on geometric features describing the contact.

One aspect, which could be particularly relevant to dissipation at the small scale is the sliding velocity relative to the contact line. 
From a continuum perspective, stress gradients are largest near a contact line so that moving parallel to a contact line would be expected to yield small dissipation and normal to it large dissipation.
Quantifying this effect is one purpose of this article.
In addition, we would like to explore how the elasticity of the solids affects dissipation.
Most solids of practical interest can be described as semi-infinite.
However, simulations assume a few layers only and keep the bottom layer fixed.
This can become problematic, when the contact radius is larger than the height of the simulated slab.
To asses the role of boundary conditions, we explore the two extreme limits of keeping the bottom most layer completely flexible or completely rigid thereby providing lower and upper bounds for the true elastic response of the substrate. 
Through this manipulation, we also effectively change the dimensionality of the objects.
A stiff bottom layer resembles a Winkler foundation, i.e., the elastic properties of the solid are that of a high-dimensional  object, while a soft foundation mimics the response of a two-dimensional sheet, each time, of course, for undulations at wavelengths clearly exceeding the height of the object.

\section{Model}

\begin{figure}[h!]
\begin{center}
\includegraphics[width=15cm]{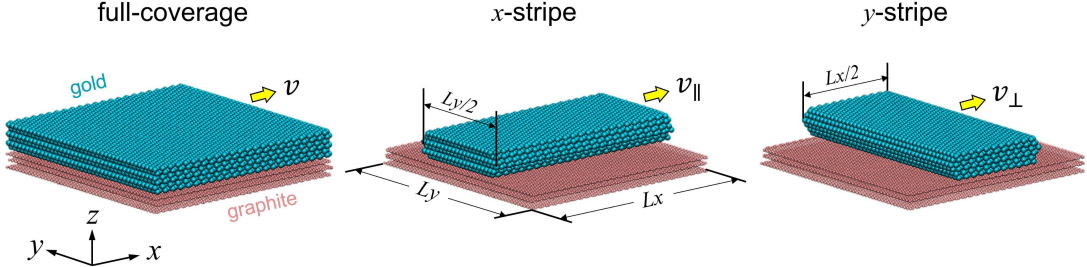}
\end{center}
\caption{{Snapshots of three simulation models at rest. Yellow arrows indicate the sliding direction, i.e., if motion is parallel ($\vert\vert$) or perpendicular ($\perp$) to the contact line. $L_x$ and $L_y$ are 11.5 and 11.6 nm, respectively.}
}\label{fig:snapshots}
\end{figure}

Using MD, we model a six-layer gold slab with (111) surface sliding against a three-layer highly oriented pyrolytic graphite (HOPG) substrate in the absence of contaminants.
Both the contact surfaces are atomically flat and defect free.
As shown in Fig.~\ref{fig:snapshots}, three  modeling scenarios are investigated, termed as: (1) full coverage, (2) $x$-stripe, and (3) $y$-stripe.
Periodic boundary conditions are applied in the $xy$-plane, so that contact lines appear only in (2) and (3).
The in-plane dimensions of the simulation cell are 11.5$\times$11.6 nm$^2$, to which the strains of the gold slab in (1) as well as those in (2) and (3) along the longitudinal directions are less than 0.3\% with respect to its minimum-energy (bulk) configuration.
The widths of the stripes in (2) and (3) are approximately one half of the corresponding cell length.

In each case, the center of mass (COM) of the HOPG's bottom layer is kept fixed while the COM of the gold top layer is constrained to move at a constant in-plane velocity ranging typically from 20$\sim$120~m/s at different angles relative to the stripe direction.
However, the COM of the gold cluster's top layer is unconstrained in the direction normal to the interface so that it moves under zero external normal load. 
Atoms in outermost layers are either constrained (``rigid'') relative to the center-of-mass of that layer or free to move (``flexible'') relative to it according to Newton's equations of motion.
From an elasticity point of view, a flexible HOPG bottom layer can be loosely associated with a freely suspended three-layer thick graphite solid. 
The elasticity of a semi-infinite graphite would be softer than the rigid bottom layer but stiffer than the flexible layer. 
The rigidity assumption for gold is made to mimic the constraining effect that an atomic-force microscope (AFM) tip moving a cluster over the surface might have on that cluster. 
Again, reality will be somewhere in between the ideal, limiting cases of perfect rigidity and flexibility. 
Neither forces nor torques or any other external constraint other than those on the COM velocities of the two outermost layers are imposed.

The interatomic interactions of graphite are described by the AIREBO potential developed by Stuart et al.~\cite{Stuart2000JCP}, while the interactions of gold are described by an EAM potential proposed by Zhou et al.~\cite{Zhou2004PRB}.
The cross interactions between the two substances are described by a Morse potential~\cite{Rosa-Abad2016RSCAdv} with a  cutoff of 10~\AA.
To improve the signal to noise ratio, thermal noise is kept small by setting the target temperature to  1~K using a Langevin thermostat.
It is only applied to the mid-graphite layer in $z$-direction with a coupling time constant of $1/\gamma = 100$~fs.
The simulation timestep is 1~fs.
All the simulations are carried out using the open-source MD code LAMMPS~\cite{Plimpton1995JCP}.

Shear stresses and (lateral) forces, whose averages are friction forces, were measured in various ways and different locations in the system, i.e., the total force acting on either the bottom-most graphite layer or the up-most gold layer as well as the interfacial forces acting between the gold and the carbon atoms.
All these forces must be identical on average during steady-state sliding, except for their sign, due to Newton's third law. 
We confirmed this to hold within the small scatter, which is due to finite run time effects. 
Averages were typically accumulated over sliding distances of 2~$\mu$m.  
In addition, the sliding-induced dissipated power $P=\gamma \sum_{i \in \mathrm{mid}}  (m \langle v_{iz}^2 \rangle -k_BT) $ adsorbed by the thermostat was averaged and also succesfully correlated with the directly measured friction forces by using the equation $P = \langle \mathbf{F} \rangle \cdot \mathbf{v}$. 
Here, $i \in \mathrm{mid}$ refers to carbon atoms in the middle graphite layer and $m$ is their mass, while $\langle ... \rangle$ indicates a time average during steady-state sliding. 

\begin{figure}[h!]
\begin{center}
\includegraphics[width=10cm]{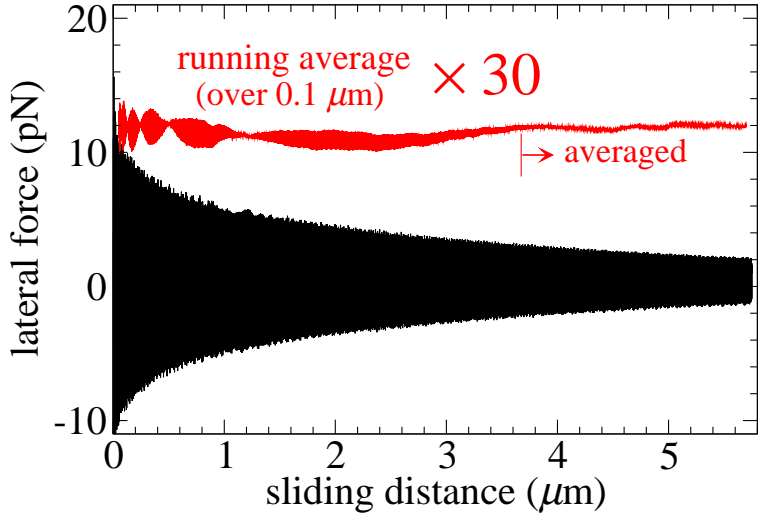}
\end{center}
\caption{{Typical dependence of the instaneous force/stress (black lines) on slid distance for the $y$-stripe at a sliding velocity of $v_\perp = 50$~m/s. Averages were only taken over the last 2~$\mu$m.}
}\label{fig:force}
\end{figure}

The shear stress was obtained by dividing the mean force in the sliding direction through the area of the adsorbate, for which each atom in the bottom gold layer was assigned the same area. 
Statistical error bars, $\Delta O^2$, of an observable $O$ are deduced from the integral over its time auto-correlation function $C_{OO}(t):=\langle O(t') O(t'+t)\rangle$ via $\Delta O^2 = \frac{2}{\tau} \int_0^\tau \! C_{OO}(t) \mathrm{d}t$, where $\tau$ is the simulated time.  
Fig.~\ref{fig:force} shows a representative measurement of the lateral force, which was conducted on a $y$-stripe sliding in $x$-direction. 
It can be seen that the mean friction force is only a small fraction of typical instantaneous forces, which thus are predominantly conservative in nature and expected for structurally lubric contacts. 

\section{Results}

\subsection{Reference cluster and preliminary considerations}
\begin{figure}[h!]
\begin{center}
\includegraphics[width=12cm]{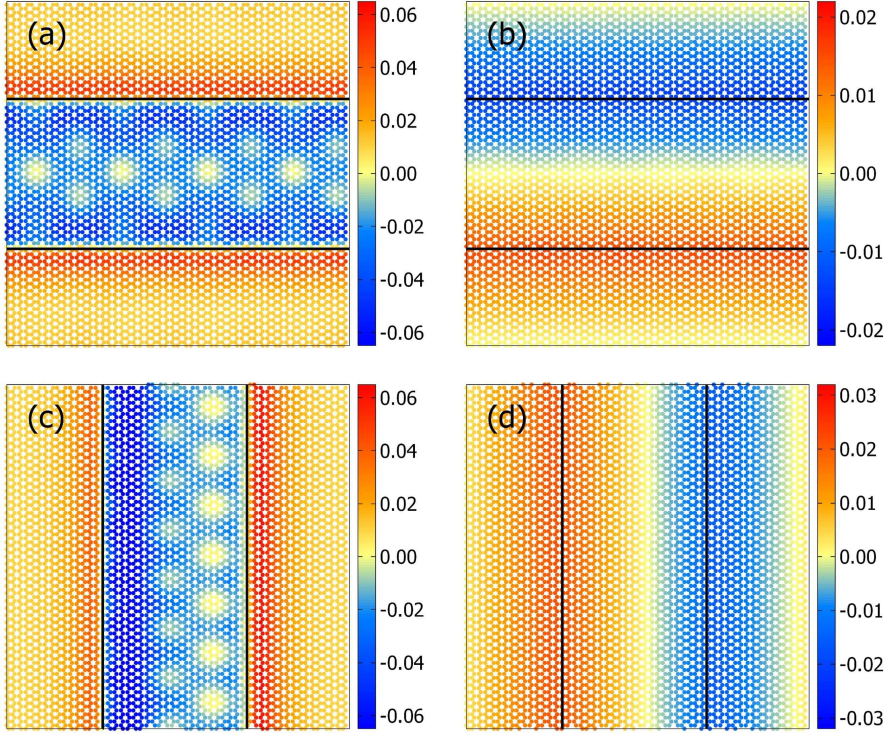}
\end{center}
\caption{\label{fig:graphiteRZ}
Static, atomic displacement in the topmost graphite layer of a $x$-stripe in (a) $z$-direction and  (b) $y$-direction relative to the equilibrium positions of a free-standing surface and similarly for a $y$-stripe in (c) $z$-direction and (d) $x$-direction.
Displacements are given in \AA.}
\end{figure}

Before discussing the results on sliding, it is useful to analyze how the interaction between ``indenter'', i.e., the cluster, and the substrate deforms the solids, because similar deformations are dragged along as the cluster is moved with respect to the substrate, whereby surface vibrations are excited, which ultimately propagate toward the bulk and get adsorbed into a heat sink, in our case into the thermostat.
Expectations from generic continuum considerations would be as follows:
The substrate can increase the interaction with the indenter by moving (near-surface) atoms below the indenter.
Such a process would increase the number of atoms below the indenter, which would make the substrate raise up below the indenter and go down right outside the contact line---assuming the Poisson's ratio of the substrate to be positive.
For our system, the situation is qualitatively different. 
Radial displacements are small, because of the stiff in-plane bonds in graphite. 
This reduces the propensity of carbon atoms to be pulled below the indenter. 
Moreover, due to the interactions being body rather than surface forces, the second graphite layer is attracted toward the gold cluster.
As a consequence, the lattice contracts below the indenter, as is evidenced in Fig.~\ref{fig:graphiteRZ}, panels (b) and (d). 
Displacements in the normal ($z$) direction turn out a factor of 3 larger than in the in-plane ($y$) direction, as revealed in Fig.~\ref{fig:graphiteRZ}(b).

\begin{figure}[hbtp]
\centering
\includegraphics[width=5.5cm]{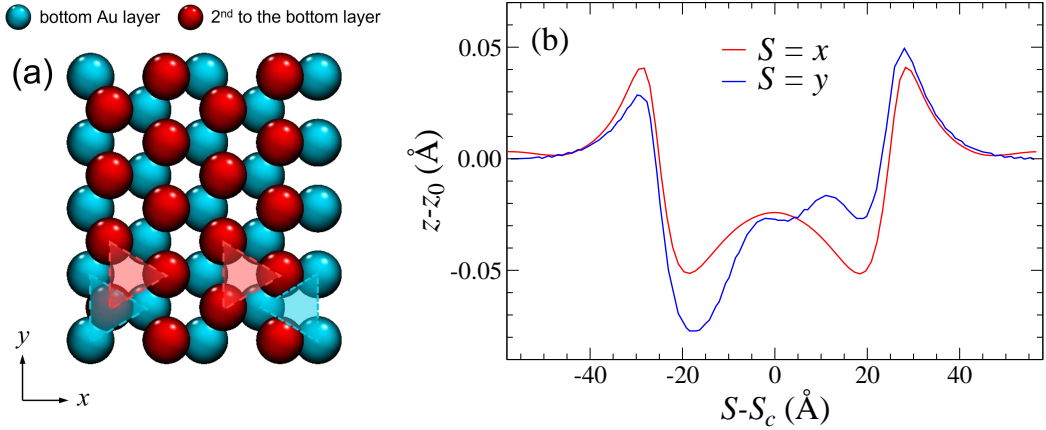}
\includegraphics[width=8.5cm]{topGrProfile.eps}
\caption{(a) Top view of two bottom-most gold layers. Cyan atoms are in direct contact with the graphite layer, while red atoms are one layer further up. (b) Spatially-resolved normal displacement of the topmost graphite layer along the finite direction of a $x$-stripe (solid, red line) and a $y$-stripe (dashed, blue line). Note that in-plane and out-of-plane direction are not true to scale. 
}
\label{fig:symmetryBreaking}
\end{figure}

Although a continuum description of the $x$-stripe and $y$-stripe would be identical, a clear difference between their normal displacements shows up.
Specifically, the $z$-displacements for the $y$-stripe reveal Moiré patterns, which clearly violate  inversion symmetry, while all other displacements obey the symmetry expectations, except, for minor  fluctuations, which are unavoidable in discrete/atomic systems.
However, the symmetry breaking for the $z$-displacements in the $y$-stripe do not necessarily reflect broken ergodicity, since the contact is not mirror symmetric about the $yz$-plane. 
In fact, it is the second gold layer, which breaks the mirror symmetry for a $y$-stripe as can be seen in Fig.~\ref{fig:symmetryBreaking}(a).
The broken synmmetry is also revealed in the normal displacements, as depicted in panel (b).

\subsection{Sliding simulations}
\begin{figure}[hbtp]
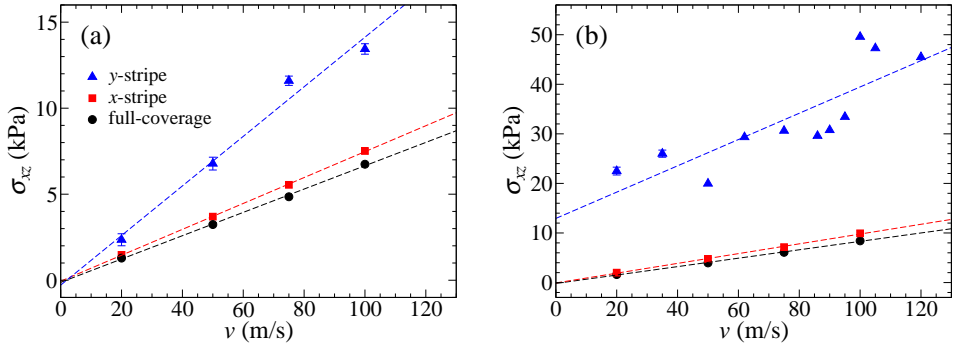

\begin{center}
\includegraphics[width=6cm]{stress_fix.eps}~~~~
\includegraphics[width=6cm]{stress_fixCOM.eps}
\end{center}
\caption{Velocity dependence of the shear stress ($\sigma_{xz}$) when sliding $x$ direction in the cases of all outermost layers being (a) rigid and (b) flexible. 
Stochastic error bars smaller than symbol size are not shown.}\label{fig:stressVel}
\end{figure}

Scaling arguments for the magnitude of instantaneous or maximum lateral forces usually focus on the contact area for amorphous clusters and on the circumference for incommensurate interfaces between crystals. 
However, this does not mean that the cluster height or the boundary conditions or constraints on the outermost layers are irrelevant.
Assuming two clusters to have identical basal planes, the thinner cluster is more compliant (or ``flexible'') than the thick one, which is effectively more ``rigid''.
This regards both in-plane and out-of-plane deformations.  
A larger in-plane compliance in the direction of sliding (i.e., in the longitudinal direction) implies increased friction, since the cluster can interlock more easily  with the counterbody.
However, an increased out-of-plane---as well as in-plane, transverse---compliance  allows the surface atoms to deflect away from irregularities on the counterface so that a larger cluster height can also increase friction.
To investigate which of the effects is more relevant for metal clusters adsorbed on graphite, Fig.~\ref{fig:stressVel} shows the shear stress as a function of velocity for (a) perfectly rigid and (b) flexible outermost layers for a $y$-stripe sliding in the $x$-direction.

Fig.~\ref{fig:stressVel}a) reveals that the motion of both stripes as well as the full layer show Stokes-like damping, roughly linear in velocity, when both outermost layers are rigid. 
The damping coefficient for the $x$ stripe roughly equals that of the full layer, while the $y$-stripe is damped a little less than twice as strongly.
The increased damping for the $y$-stripe can be easily rationalized, since its velocity is perpendicular to the contact line. 
From a continuum perspective of brittle fracture, a (small) hysteresis between a closing crack at the leading edge and an opening crack at the trailing edge must be expected, which adds to the bulk or areal dissipation far away from the crack~\cite{Popov2021F}. 

\begin{figure}[hbtp]
\begin{center}
\includegraphics[width=8cm]{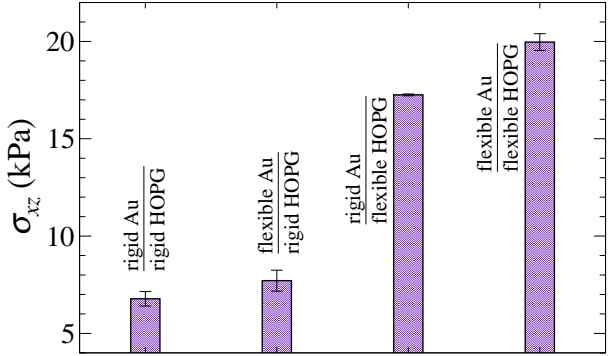}
\end{center}
\caption{Comparison of mean shear stresses for a $y$-stripe moving with $v_\perp = 50$~m/s under different boundary conditions.}\label{fig:chartBars}
\end{figure}

After releasing the rigidity constraint at both outermost layers, the lateral force opposing the motion of both  $x$-stripe and  full layer parallel to $x$ still is Stokesian with a marginally,  $\mathcal{O}(10\%)$, increased damping coefficient.
Interestingly, the $y$-stripe reveals a much increased resistance to sliding and a rather weak velocity dependence of the kinetic friction on sliding velocity at small $v$.
This increased friction is related predominantly to the increased compliance of the flexible graphite substrate, as can be seen from Fig.~\ref{fig:chartBars}.

\begin{figure}[hbtp]
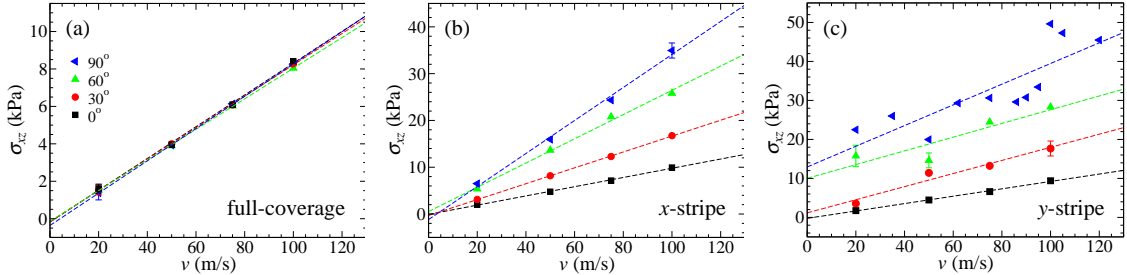

\begin{center}
\includegraphics[width=4.8cm]{stress_full_angle.eps}~
\includegraphics[width=4.8cm]{stress_xstripe_angle.eps}~
\includegraphics[width=4.8cm]{stress_ystripe_angle.eps}
\end{center}
\caption{Dependence of shear-stress on sliding velocity for (a) full-coverage, (b) $x$-stripe, and (c) $y$-stripe using different sliding directions ranging from parallel ($0^\circ$) to perpendicular ($90^\circ$) to the contact line. Both outermost layers are flexible in all cases. }\label{fig:stressOrient}
\end{figure}

Fig.~\ref{fig:stressOrient}  highlights the interplay of sliding direction and mean lateral force in the case of flexible outermost layers, which we find to be the most interesting case, since it can deviate from Stokes damping.
The full-coverage layer does not show any detectable direction dependence.
The $x$-stripe reveals a damping coefficient, which increases with increasing angle between contact line and sliding direction, while the $y$ stripe transitions between Stokes and Coulomb-like friction at an angle between 30$^\circ$ and 60$^\circ$. 

\begin{figure}[hbtp]
\centering
\includegraphics[width=15cm]{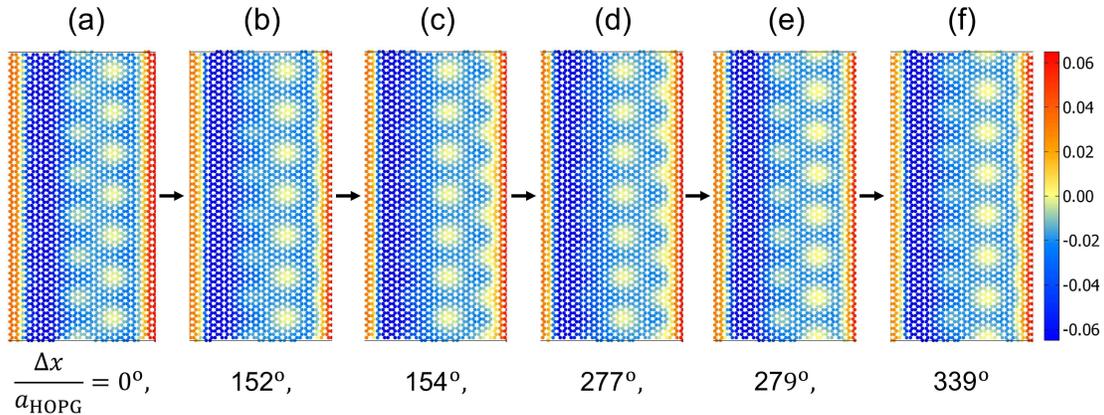}
\caption{Normal displacements after different sliding distances, in units of $\Delta x / a_\textrm{HOPG}$: (a) 0, (b) 0.422, (c) 0.428, (d) 0.769, (e) 0.775, and (f) 0.942.}
\label{fig:movie_figure}
\end{figure}

Instabilities are required, in order for Coulomb-like friction to occur~\cite{Prandtl28,Tomlinson1929,Muser2002PRL}.
Movies of the displacements were produced to characterize the instabilities, i.e., animated versions of all panels shown in Fig.~\ref{fig:graphiteRZ}. 
They clearly reveal that the normal displacement of the $y$-stripe moving in $x$-direction displays quasi-discontinuous dynamics, while all other displacement fields evolve continuously with time. 
Snapshots of these movies are shown in Fig.~\ref{fig:movie_figure}, from where it becomes apparent that the observed Moiré patterns only changed marginally in the relatively large time periods separating panels (a) from (b), (c) from (d), and (e) from (f), but quite substantially during the brief time periods separating configuration (b) from (c) and (d) from (e). 

The dynamics loosely resemble that of a crack front, which discontinuously advances by a distance of the order of one atomic spacing~\cite{Wang2020EPL}.
However, for the studied case, different types of instabilities occur, where the corrugated pattern does not only advance by a small multiple-integer of the graphite lattice constant but simultaneously, e.g., in the transition from Fig.~\ref{fig:movie_figure} panel (b) to (c), but also undergoes a phase shift of 180$^\circ$ from (d) to (e).
As is always the case for Coulomb friction, the energy dissipated in the process is approximately the difference of the closest energy minimum of the configuration just before and just after the pop.
It is relatively insensitive to the rate with which locally released energy or heat is transported away from the interface, unless the heat conductance is so small that the interface heats up substantially. 

\begin{figure}[hbtp]
\begin{center}
\includegraphics[width=8cm]{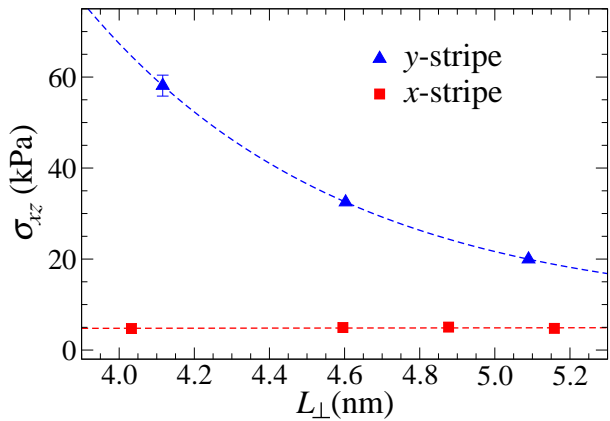}
\end{center}
\caption{
Area dependence of shear stress for the $x$ and $y$-stripe geometries and flexible outermost layers at $v_\perp = 50$~m/s. Here, the size of the substrate was kept constant and the length of the slab varied parallel to the sliding velocity direction. The blue dashed line shows an exponential dependence of the shear stress with $L_\perp$ with $\sigma(L_\perp \to \infty)$ constrained to the shear stress of the shear-stress obtained at full coverage ($\approx 5$~kPa). 
}\label{fig:stress_area}
\end{figure}

Finally, the size-dependence of the Coulomb shear stress is investigated in Fig.~\ref{fig:stress_area}.
It decreases with the length of the $y$-slab much more quickly than simply with $1/L_x$.
Specifically, the decay is consistent with an exponential dependence  on the relatively small investigated domain. 
This means that the contribution of the near-contact line zone cannot be simply assigned a unique value, but it appears to be a non-local contribution. 

\section{Conclusions}

In this paper, we studied the kinetic friction of adsorbed, periodically continued gold stripes, sliding past a graphite substrate under their mutual adhesive attraction.
We found the systems to exhibit structural lubricity for the most part in that mean lateral forces disappeared quickly, i.e., linearly with velocity. 
When atoms in the outermost layers of the graphite substrate and the gold adsorbate were constrained to zero displacement relative to their equilibrium position in the respective outermost layer, Stokes friction was observed in all cases.
When sliding a full-coverage cluster kn any direction or a finite-width slab parallel to the direction of the stripe, Stokes stress, defined as Stokes friction per unit area, turned out roughly $\sigma_{20\rm-m/s}=$1.3~kPa at a sliding velocity of 20~m/s.
This is the same as the shear stress that would be obtained with a $\sim 1.5$~$\mu$m thick film of lubricant as water at ambient conditions, $\eta = 0.1$~mPa$\cdot$s, and stick boundary conditions. 
In these ``Stokesian cases'' of our study, dissipation can be related to quasi-elastic deformations of the two solids in contact and the unavoidable hysteresis, which is consistent with linear harmonic theory, as  proposed, for example, in the pioneering study by Adelman and Doll~\cite{Adelman1976JCP} or the more modern formulation by Kajita et al.~\cite{Kajita2010PRB} using time-dependent elastic Green's functions. 

Allowing the outermost layers to be flexible increased the friction in all cases, but most so for the $y$-stripe when sliding parallel to the armchair direction of the graphite lattice, in which case Coulomb friction occurred. 
Although the shear stress also increased for the $x$-stripe, in fact by a factor of four when sliding perpendicular rather than parallel to the stripe, it remained Stokesian. 
This latter behavior can be rationalized to be a consequence of a slowly-moving adhesive crack pair, in which the crack closure on the leading edge does not fully recuperate the energy needed to open the crack at the closing edge.
When elastic or phonon relaxation times are small, this process leads to a hysteresis linear rather than algebraic in velocity~\cite{Muser2022EPL}. 
When moving parallel to the slab, no contact-line contribution was observed, which, however, can be due to the fact that both solids had a very smooth, unjagged edge.

For flexible outermost layers, which in the case of the modeled graphite  can be associated with a suspended substrate, Coulomb friction was induced by the added compliance.
This result is in agreement with the pioneering study of Lee et al.~\cite{Lee2010Science} on the friction of suspended layered solids, including graphite, where friction was observed to continuously decrease with width, which is in line with the argument that long-range elastic restoring forces are a key element for superlubricity~\cite{Shinjo1993SS,Muser2004EPL}.
Lee et al.~\cite{Lee2010Science}  related the increased friction to the easier out-of-plane puckering for thin sheets.
We have little reason to object to that conclusion, but refine the mechanism for our system, in that puckering occurs not only in front of the indenter but also below it.
Moreover, the induced undulations below the indenter near the contact line have Moiré pattern characteristics and move continuously for the most part with occasional quasi-discontinuities at isolated moments of time. 
These Moiré patterns differ from those considered earlier in the context of friction, see, e.g., 
Fig.~1 
in Ref.~\cite{He1999S} or the Moiré pattern analysis by M\"user~\cite{Muser2001PRL} in that normal rather than in-plane displacements form the Moiré pattern. 
Placing our results into a historical context, our Moiré pattern instabilities are rather a Fourier version of the discontinuous dynamics suggested by Tomlinson~\cite{Tomlinson1929}, who considered instabilities normal to the interface, rather than those by Prandtl~\cite{Prandtl28}, who focused implicitly more on motion within of atoms within their planes. 

Despite some commonality with experiments, the shear stresses in our system are extremely small, i.e., about 5~kPa when the linear size of the stripe extends over 25~nm, which is roughly one order of magnitude less than typical peaks in lateral forces or than static friction. 
In most other situations, instabilities are required to lead to detectable friction, i.e., the friction in our systems would still have to be classified as superlubric by all means, yet, we would not label them as structurally lubric, due to the presence of multi-stability at scales smaller than the width of the stripes. 
While it remains to be seen to what extent our puckering-Moiré-pattern mechanism matters in real (UHV) systems, our study certainly revealed that utmost care has to be taken when modeling the friction of graphite layers. 
It also is a paradigm support for Coulomb's ingenious insight of the origin of friction, reiterated in the very beginning of our introduction, except that in our case the roughness that needs to be overcome to continue sliding is not pre-existing but the consequence of the coherence that graphite and gold want to form due to their proximity.

\section*{Conflict of Interest Statement}
The authors declare no conflict of interest.

\section*{Author Contributions}
HG set up, run, and analyzed the simulations and edited the paper. 
MHM designed the study, assisted with the analysis of the simulations, and wrote the paper.

\section*{Funding}
This research was founded by the German Research Foundation (DFG) through grant number GA 3059/2-1.

\section*{Acknowledgments}
We thank Mehmet Baykara for many stimulating discussions. 
%


\bibliographystyle{abbrv}
\bibliography{reference}

\end{document}